\documentclass[12pt,preprint]{aastex6}
\usepackage{natbib,epsfig}

\begin{document}

\slugcomment{Submitted to the Astrophysical Journal, April 2017}
\title{Constraints on the Growth and Spin of the Supermassive Black Hole in M32 From High Cadence Visible Light Observations}

\author{R. Chary\altaffilmark{1}, G. Hallinan\altaffilmark{1}, L. K. Harding\altaffilmark{2}, N. S. Saini\altaffilmark{1},  H. E. Schlichting\altaffilmark{3}}

\altaffiltext{1}{Division of Physics, Math \& Astronomy, California Institute of Technology, Pasadena, CA 91125; rchary@caltech.edu}
\altaffiltext{2}{NASA/Jet Propulsion Laboratory, Pasadena, CA 91109}
\altaffiltext{3}{University of California, Los Angeles, CA 90024 \& Massachusetts Institute of Technology, Cambridge, MA 02139}

\begin{abstract}
We present 1\,s cadence observations of M32 (NGC221) with the CHIMERA instrument 
at the Hale 200-inch telescope of the Palomar Observatory. 
Using field stars as a baseline for relative photometry, we are able to construct
a light curve of the nucleus
in the $g'$ and $r'$ band with
1$\sigma$=36 milli-mag photometric stability. We derive a temporal power spectrum for the nucleus and find no evidence
for a time-variable signal above the noise as would be expected if the nuclear black hole were accreting gas. Thus, we are unable to
constrain the spin of the black hole although future work will use this powerful instrument to target more actively accreting black holes.
Given the black hole mass of 2.5$\pm0.5\times$10$^{6}$\,M$_{\sun}$
inferred from stellar kinematics, the absence of a contribution from a nuclear time-variable signal places an upper limit on
the accretion rate which is $4.6\times10^{-8}$ of the Eddington rate, a factor of two more
stringent than past upper limits from {\it HST}. The low mass of the black hole despite the high stellar density suggests 
that the gas liberated by
stellar interactions was primarily at early cosmic times when the low-mass black hole had a small Eddington luminosity.
This is at least partly driven by a top-heavy stellar initial mass function at early cosmic times which is an efficient producer of stellar mass black holes.
The implication is that supermassive black holes likely arise from seeds formed through the coalescence of 3-100\,M$_{\sun}$ mass black holes that then accrete gas produced through stellar interaction processes.
\end{abstract}

\section{Introduction}

The kinematics of stars and gas in the nuclei of massive galaxies have provided strong evidence for
the presence of supermassive ($>10^{5}$\,M$_{\sun}$) black holes at their centers \citep{2000ApJ...539L...9F, 2000ApJ...539L..13G}. Although the origin
and growth of these supermassive black holes is unclear, it is thought that they likely accreted
significant amounts of gas over the galaxy's lifetime, undergoing sporadic and luminous phases of rapid growth.

Black holes are characterized by three fundamental properties, mass, spin and electrical charge. The charge is thought
to be zero in astrophysical black holes; the mass has been determined through kinematics of the surrounding stars and gas
\citep{1995Natur.373..127M, 1997MNRAS.291..219G, 1998ApJ...509..678G} but little is known about their spin.  
Although the spins of stellar mass black holes in binary systems have now been constrained through the detection of gravitational
waves \citep{2016PhRvX...6d1014A} and the 
X-ray reflection spectrum \citep{2014ApJ...784L...2K, 2016ApJ...826L..12E}, these 
measurements are dominated by the end stages of the dynamical evolution of the system.
The inherent spin of a supermassive black hole is a challenging quantity to observe but constraints have been
obtained through their X-ray reflection spectrum, particularly the gravitational redshift of iron lines which appear to arise at distances of tens of Schwarzschild radii
\citep{2007ARA&A..45..441M, 2011ApJ...736..103B, 2011MNRAS.411.2353P}. 
The observations seem to confirm the hypothesis that
these supermassive black holes have significant angular momentum associated with them, approaching the maximally
spinning Kerr metric
\citep{Volonteri2005}. However, it is unclear if there is a bias in the measurement whereby the most X-ray luminous black holes are presumably the ones
undergoing large accretion events, which in turn have spun up the black holes \citep{1974ApJ...191..507T}. Measurement of the spin
of a relatively quiescent black hole and its relationship with the kinematics of the galaxy could shed light on the formation mechanism for the supermassive black hole,
revealing whether the primary mechanism of growth was through accretion or through merging of comparable mass black holes which may be low
frequency gravitational wave sources detectable by a future gravitational-wave interferometer like Laser Interformeter Space Antenna (LISA).

If a black hole is not spinning, then the radius
of the innermost stable circular orbit (ISCO) of a radiated photon from a particle accreting onto the black hole is 3 times the Schwarzschild radius of the black hole.
On the other hand if a spinning black hole is accreting gas, the radius of the ISCO is
just beyond the Schwarzschild radius with the radius depending on the angular momentum of 
the black hole \citep{1970Natur.226...64B, 1972ApJ...178..347B, 1974ApJ...191..507T}. In the standard accretion picture, a disk of infalling material would be an optically thick,
geometrically thin blackbody of several 10$^{4}$\,K responsible for optical/UV emission which is Comptonized by a hot corona to emit X-rays.
By measuring the shortest time scale variations in the luminosity of an accreting black hole whose mass is known,
we would be able to probe the timescales of variations
in the accretion disk. By using the light travel time, we can derive the radius of the disk corresponding to the last stable
orbit and thereby place constraints on the spin of a black hole. Specifically, if $t_{min}$ is the smallest timescale for detected variability,
then, the radius of the ISCO ($r$) in units of the Schwarzschild radius of the black hole is: 
\begin{eqnarray}
r & = & \frac{c^{3}\times t_{min}}{4GM}\\
a & = & \mp\left( \frac{r^{3/2} - 3Mr^{1/2}}{2M^{1/2}} \right )
\end{eqnarray}
and $a$ is the spin of the black hole in units of $J/cM$ where $J$ is the angular momentum of the black hole, $M$ is the mass in solar units and
$G$ is the gravitational constant. Thus, $a=0$ is a Schwarzschild black hole and $a=1$ is a maximally spinning Kerr black hole \citep[See][for details]{1972ApJ...178..347B}.

Here, we present 1s cadence observations made with the Caltech high-speed multi-color camera (CHIMERA), a new instrument at the Hale 200-inch \citep{2016MNRAS.457.3036H} with the goal of measuring
the light curve of a nearby galaxy which is thought to have a supermassive black hole in its nucleus \citep{Tonry1984, Lauer1992}. M32 was chosen since it is bright (U$\sim$9 mag), nearby (0.8 Mpc)
and an early type galaxy whose nuclear
light appears to have no apparent contribution from an AGN even at the spatial resolution 
of {\it Hubble} \citep{Lauer1992}. Yet, due to the cuspiness of the stellar surface brightness profile, the black hole must be accreting
stars, albeit at an uncertain rate. If the stellar accretion is episodic, then the black hole may become luminous for only intervals of time
although the latest timescales for tidal disruption events in the vicinity of black holes suggest accretion times of 100s of days.
Furthermore, its estimated black hole mass of 2.5$\pm0.5\times$10$^{6}$\,M$_{\sun}$
\citep{2002MNRAS.335..517V} is small enough that
density variations near the last stable orbit in any surrounding accretion disk would be observable on $\sim40-120$\,s time scales. By analyzing the light curve
of the nucleus, we are able to place constraints on the amount of variability in the optical bands and the preferential timescale for variation
which if detected, would constrain the spin of a relatively inactive black hole.

\section{Observations and Data Reductions}

The data on M32 ($\alpha$,$\delta$)=(00h42m41.832s +40d51m55.03s in J2000)
were taken with CHIMERA simultaneously in the Sloan $g'$ and $r'$ bands. 
CHIMERA is located at the prime focus of the 200-inch Hale telescope at the Palomar Observatory and subtends a field of view of
5$\arcmin\times5\arcmin$ with a pixel scale of 0.28$\arcsec$/pixel. The data were taken with exposure times of 1s
on 2016 Dec 4th between 0457 and 0545 hrs and Dec 6th between 0320 and 0450 hrs UTC.  26 sets of 100 exposures
were taken on the first night and 49 sets of 100 exposures on the second night with gaps of 5\,s between each set.
The nights were photometric. However, the seeing in the optical bands
was varying between 2.5 and 4.5$\arcsec$ (median 3.5$\arcsec$; full width at half maximum; FWHM) on the first night and between 1-2.5$\arcsec$ (median of 1.3$\arcsec$
FWHM) on the second night. The position of the nucleus of the galaxy on the detector was allowed to drift over the duration of these observations by about 9$\arcsec$. 

After bias subtraction and flat fielding, aperture photometry was performed on the nucleus of the galaxy as well as four nearby
stars which are within the field of view (Table 1). There are several issues with obtaining precision light curves from seeing-limited data. First is the fact that due to point spread function (PSF) variations, the
fraction of source flux falling within the aperture is changing with time; this can be corrected since the fraction of light falling outside the aperture should be the same for all unresolved sources. 
Second, seeing causes different amount of bulge light to scatter within the extraction aperture used for photometry of the nucleus. The only way to account for this effect when
the PSF is varying is to measure the temporal power spectrum of the seeing variation and ensure that it does not correspond to a similar temporal variation in the photometry.
Although fitting surface brightness profiles to the bulge light is feasible, these are dominated by systematics at the percent level even in space based data where the PSF is constant
\citep[see e.g.][]{Lauer1992}.
A third effect has to do with variation in the sky estimation because of seeing; when the field of view is filled with a population of unresolved stars, seeing causes
different amounts of light to enter the sky aperture which in turn modulates the sky background subtraction. 

As a result, photometry was performed in three ways: 1) two-dimensional Gaussian fits to the sources with the sky background local to the source as a free parameter;
2) in a circular aperture of a fixed radius of 1.8$\arcsec$ 
and 3) in a circular aperture whose radius corresponds to the seeing FWHM. 
Sky subtraction is challenging since the galaxy effectively fills the field of view. The sky was measured by both adopting a single sky per frame which is the median of the sky within a large annulus
and by using a local sky which is 9$\arcsec$ away from each of the reference stars and 60$\arcsec$ away in the case of the nucleus of the galaxy. 
It was found that using a local sky and fixed constant aperture resulted in the smallest variation in photometry (Table 2) and our best light curves are based on this method.

The first night of data was rejected since the standard deviation of the brightness of the reference stars was almost a factor of two worse than the second night
where the seeing was much better. The raw light curves
in the $g'-$ band from the second night is shown in Figure 1 with the table of stars used for relative photometry listed in Table 1. The calibrated light curve was derived after
measuring the instantaneous zero point on Star 2 with respect to the {\it Gaia} G-band \citep{Gaia2016} and using it to calibrate the photometry of each of the other reference stars and the nucleus of M32. It is
shown in Figure 2a and 2b for the $g'-$ and $r'-$bands respectively. The standard deviation of the resultant calibrated photometry of the stellar sources is 10 milli-mag in the $g'-$band and 8 milli-mag in the
$r'-$band at 12th mag. At 16th mag, we are able to obtain about 23 milli-mag in the blue and in the red bands (Table 2). When we limit the data to only the highest quality seeing
i.e. less than 1.5$\arcsec$, the corresponding variation in the photometry of the M32 nucleus is 1$\sigma$=36 millimags, the difference from a reference star of similar brightness
being almost entirely due to starlight
from the spheroid entering the photometric extraction aperture. Future work with data taken in better seeing will attempt to model this observational effect in greater detail.

After calibration of the photometry as described above, we standardize the time baselines. There are intervals of $\sim$5s between sets of 100$\times$1.102s frames and there is the overhead of the frame transfer time of 1.2ms per frame as well. So the start times of subsequent frames relative to the first frame are 1.103, 2.207, 3.310,\ldots109.231, 114.11, 115.21\,s etc. 
As shown in Figure 2, there are intervals of time when the brightness of the nucleus increases relative to the reference stars in the calibrated light curve. This is
highly correlated with worsening seeing and is due to stellar light from the galaxy being scattered into the extraction aperture which is located on the nucleus. 
We therefore reject all photometry where the seeing is worse than 1.5$\arcsec$ and rescale the light curves of Star 1 and Star 3 by the ratio of the brightness 
between M32 and the stars (Table 1 and Figure 2). Star 1 is fainter than M32 by 0.9 mag while Star 3 is fainter by 4.86 mag. As a result, the light curve of
Star 3 is much noisier (Figure 2). 
We then
take the Fourier transform of the masked, calibrated light curves to obtain an un-binned temporal power spectrum for both M32 and the two scaled reference stars.
We then difference the temporal power spectrum of M32 from the power spectrum of the scaled Star 1; we also took the difference in the power spectrum
between the scaled Star 1 and Star 3 as a baseline.  
Both power spectra were binned in intervals of 5 cycles in frequency space. The uncertainty in the power spectrum was measured as:
\begin{equation}
\sigma_{Power}^2 = \sigma_{|FFT_{AGN}|-|FFT_{Star1}|}^2 +  \left(\frac{\sigma_{|FFT_{Star3}|-|FFT_{Star1}|}}{1.8}\right)^2
\end{equation} 
where the factor of 1.8 is to account for the larger scatter in the photometry of a fainter star as described earlier.
We then searched for any statistically significant excess power in the temporal power spectrum of M32.
Having four reference stars in the field allows us to calibrate out systematics in the light curve due to any part of the observing or instrument.

\section{Analysis}

Figure 3 shows the region in frequency space in the temporal power spectrum corresponding to a maximally spinning and Schwarzschild black hole. 
We find no evidence for excess power in the power spectrum in this regime with the 3$\sigma$ upper limit on any periodic signal being 2.8$\times$10$^{-5}$\,Jy. 
The only part of the entire frequency space in which there is some 
evidence for excess power is on time scales of 1.8-2.2\,s. We conclude this is because of stochastic seeing variations since the power
spectrum of the seeing shows similar peaks on the $<3$\,s timescale.

We also searched a contiguous interval of 200\,s in the data when the standard deviation in the seeing was the smallest and assessed if there was any excess in the power spectrum in that data. We found no evidence of any statistically significant excess. 
We conclude therefore that the contribution to the optical photometry from accretion onto the supermassive black hole is negligibly small, of order 0.03\%. 
This is consistent with the analysis of \citet{Lauer1992} who find no variation in optical colors as a function of distance from the nucleus in {\it HST} data
indicating that the nuclear brightness is dominated by a cusp in the stellar light distribution and has no contribution from accretion onto the black hole.

Similarly, we do not find evidence for any statistically significant excess in the power-spectrum in the $r'-$band
either. That is not entirely surprising since in the standard accretion disk scenario, the ratio in brightness between the nucleus of M32
and the surrounding starlight is thought to increase as one goes to shorter wavelengths. The absence of any excess in the power spectrum in the $g'-$band
thereby would argue against the presence of an excess in the $r'-$band, consistent with this assumption. 

We therefore conclude that the SMBH in M32 is indeed quiescent and gas accretion either from tidal disruption events or from stellar feedback processes
is not occuring. Indeed, it has been estimated that the rates of TDEs in galaxies
with similar black hole masses are $\sim1-3\times10^{-4}$\,yr$^{-1}$ \citep{Metzger}. The upper limit of 2.8$\times$10$^{-5}$ Jy to the total nuclear photometry
of 0.15 Jy, corresponds to a 3$\sigma$ $g'-$band luminosity limit of 1.47$\times10^{37}$\,erg\,s$^{-1}$; this is more than a factor of two fainter than the limits derived from
the X-rays or space-based optical data and corresponds to $4.6\times10^{-8}$ of the Eddington luminosity of the black hole. 

Based on our derived upper limit on the contribution of the SMBH to the nuclear photometry and past measurements of the surface brightness slope ($\Sigma\sim r^{-1/2}$)
and  the stellar density in the nucleus of M32 \citep{Lauer1992},
we can place interesting constraints on the evolution of the nucleus. 
The dynamical evolution of a dense stellar cusp surrounding a black hole and the resultant luminosity evolution of the black hole has been modeled in some detail \citep{1991ApJ...370...60M, 2006ApJ...649...91F}. The various physical processes responsible for driving the luminosity evolution of the black hole are tidal disruption events, stellar collisions
and gas infall from stellar feedback processes.  
Although the peak luminosity of a black hole in a galaxy with a central density of $>3\times10^{7}$\,M$_{\sun}$\,pc$^{-3}$ is thought to be in the range $\gtrsim 2\times10^{44}$\,erg\,s$^{-1}$, the temporal
change in its luminosity is thought to have a long time constant, $\sim t^{-3}$ \citep{1991ApJ...370...60M} arising from gas infall due to stellar feedback processes.
For example, \citet{1991ApJ...370...60M} estimate that even after a Hubble time, in a radiatively inefficient scenario, 
the luminosity of a black hole should be 10$^{-6}$ of its peak luminosity and results in black holes which are much more massive than the black hole in M32. 
Specifically, their simulations indicate that dense stellar cusps should produce black holes with mass $>10^{8}M_{\sun}$.
Our non-detection and the relatively low black hole mass 
imply that mass accretion of gas from stellar collisions and feedback processes is less efficient than even the least efficient scenario
modeled, such that only $\lesssim$2.5\% of that mass is accreted onto the black hole. 

The origin of this low efficiency is puzzling. One possibility is that most of the gas liberated from the interaction of the stars occurs early in the life of the black hole
where its mass and Eddington luminosity are small. This would require the stellar IMF in the galaxy at early cosmic times to be top-heavy (dN/dM$\propto M^{-1.7}$), 
similar to that found
in some compact, high-redshift galaxies \citep{2008ApJ...680...32C, Shim2011}. In this scenario, four times more of the central stellar mass density would be in high mass stars which lose their gas
efficiently through stellar interactions and winds and collapse into stellar black holes on timescales of the massive star lifetime i.e. $\sim10^{7}$\,yr.
In comparison, a black hole of mass 2.5$\times10^{6}$\,M$_{\sun}$  is built up on timescales of $\gtrsim$0.5\,Gyr if it grows through Eddington limited accretion onto a stellar mass black hole seed (i.e. $3-100$\,M$_{\sun}$).
Due to the relatively small mass of the black hole on timescales corresponding to gas ejection, only a small fraction of the gas is accreted onto it, resulting in a relatively low-mass black hole in a dense stellar cluster.
Furthermore, in this scenario, the early stages of the nuclear black hole growth would be dominated by preferential mass segregation of the stellar black holes in the nuclear regions followed by dynamical friction resulting in them coalescing into a $\sim$1000\,M$_{\sun}$ seed supermassive black hole \citep{2006ApJ...649...91F}. Evidence favoring this hypothesis comes from an analysis of the nuclear stellar populations by \citet[][and references therein]{2009MNRAS.396..624C} who find strong evidence
that the nucleus has at least half its light arising from a low-metallicity, evolved stellar population of age $\sim$10\,Gyr. However, they argue that interpretation of the metal yields is extremely uncertain
and model dependent and thereby cannot be leveraged as a tracer of the early IMF.

Alternately, while it is plausible that the duty cycle of gas accretion onto the black hole is very small $\lesssim$2.5\% compared to the calculations of \citet{1991ApJ...370...60M}, it is challenging to understand why in a dense stellar cluster with ongoing stellar interactions and presumably
stellar winds, the mass accretion rate onto the black hole is so strongly suppressed. For example, \citet{2006ApJ...649...91F}  adopt that 6.5\% of the gas emitted by stellar evolution is accreted
by the black hole and still have a stellar cusp which is an order of magnitude less dense than that observed in M32 for its black hole mass. Thus, to account for the low mass of the black hole and high core mass density in stars,
some other mechanism would need to be at play to suppress infall of gas emitted by stellar interactions.

\section{Conclusions} 

We have presented a technique which could reveal the spins of supermassive black holes using temporal brightness variations in the optical bands. These
variations may arise due to accretion of material from a tidal disruption event in the nucleus or due to accretion of the interstellar medium
through an accretion disk. As a proof of concept, we targeted the quiescent super-massive black hole in M32 and find
no evidence for variability in the nucleus and also no contribution from accretion onto the supermassive black hole. We place stringent upper limits on the 
optical luminosity of the nuclear black hole, a factor of two stronger than previous work. Given models for the dynamical evolution of stars in the vicinity of the
black hole, the absence of any signal indicates that gas accretion by the black hole from the central stellar cusp is remarkably inefficient. This is surprising and suggests that high density stellar cusps may unbind a large fraction of their gas at early cosmic times when the black hole is of low mass, limiting its accretion rate. This would further imply that the seeds of super massive black hole formation form
by mass segregation and coalescence of stellar mass black holes followed by accretion of the gas that is produced by stellar interactions and winds.
Furthermore, previous measurements of spins of SMBH have used X-ray
reflection spectra which necessarily imply that the mass accretion rate of the black hole is large and that the spins may be biased high by recent accretion.
Thus, this technique could complement X-ray derived spins in objects where the mass accretion rate of the black hole is relatively small.
Future work will utilize the unique capabilities of CHIMERA to target SMBH with spins derived from the X-ray data to verify the potential of this technique
and assess if X-ray derived spins are representative of black hole spins.

\begin{figure}[htbp]
\epsscale{0.7}
\begin{center}
\plotone{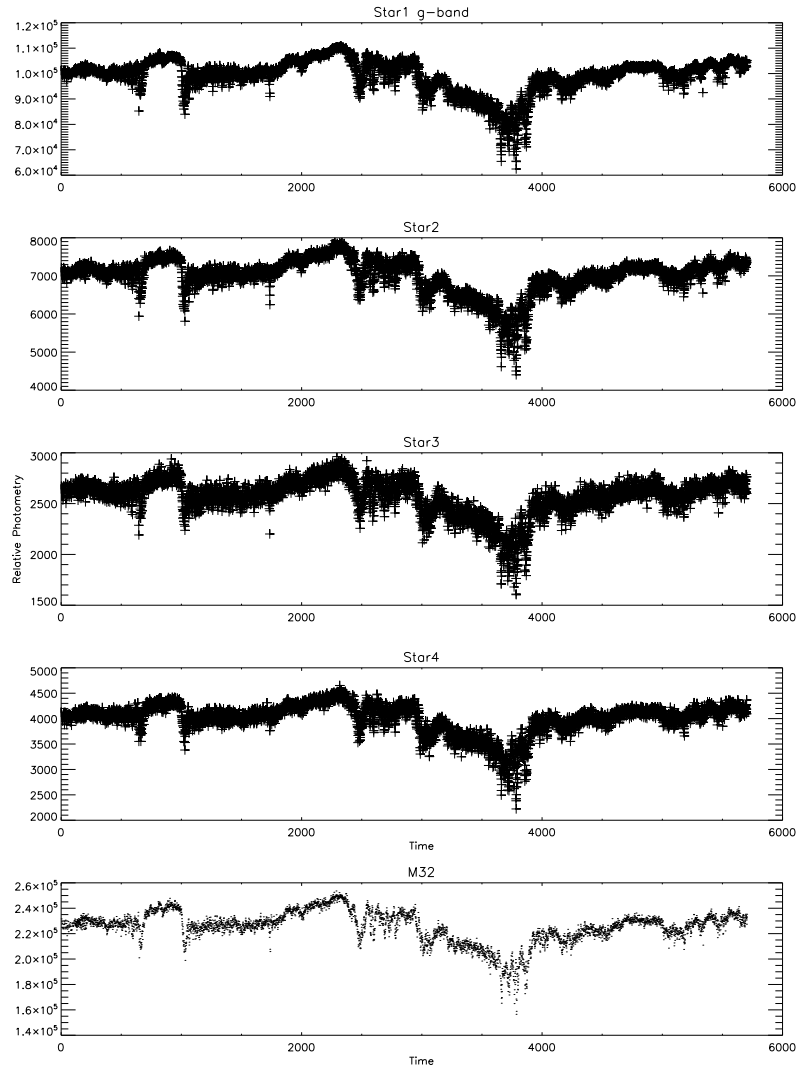}
\caption{Raw $g'-$band light curve of four reference stars and the nucleus of M32 with no corrections applied.
The abscissa shows sample number which is a proxy for time in seconds while the ordinate shows relative photometry in counts. Clearly
photometric variations can be significant, as much as a factor of 2. However, the variations are clearly correlated between the reference sources
and the target and can be corrected.}
\label{fig1}
\end{center}
\end{figure}

\begin{figure}[htbp]
\epsscale{1.15}
\begin{center}
\plottwo{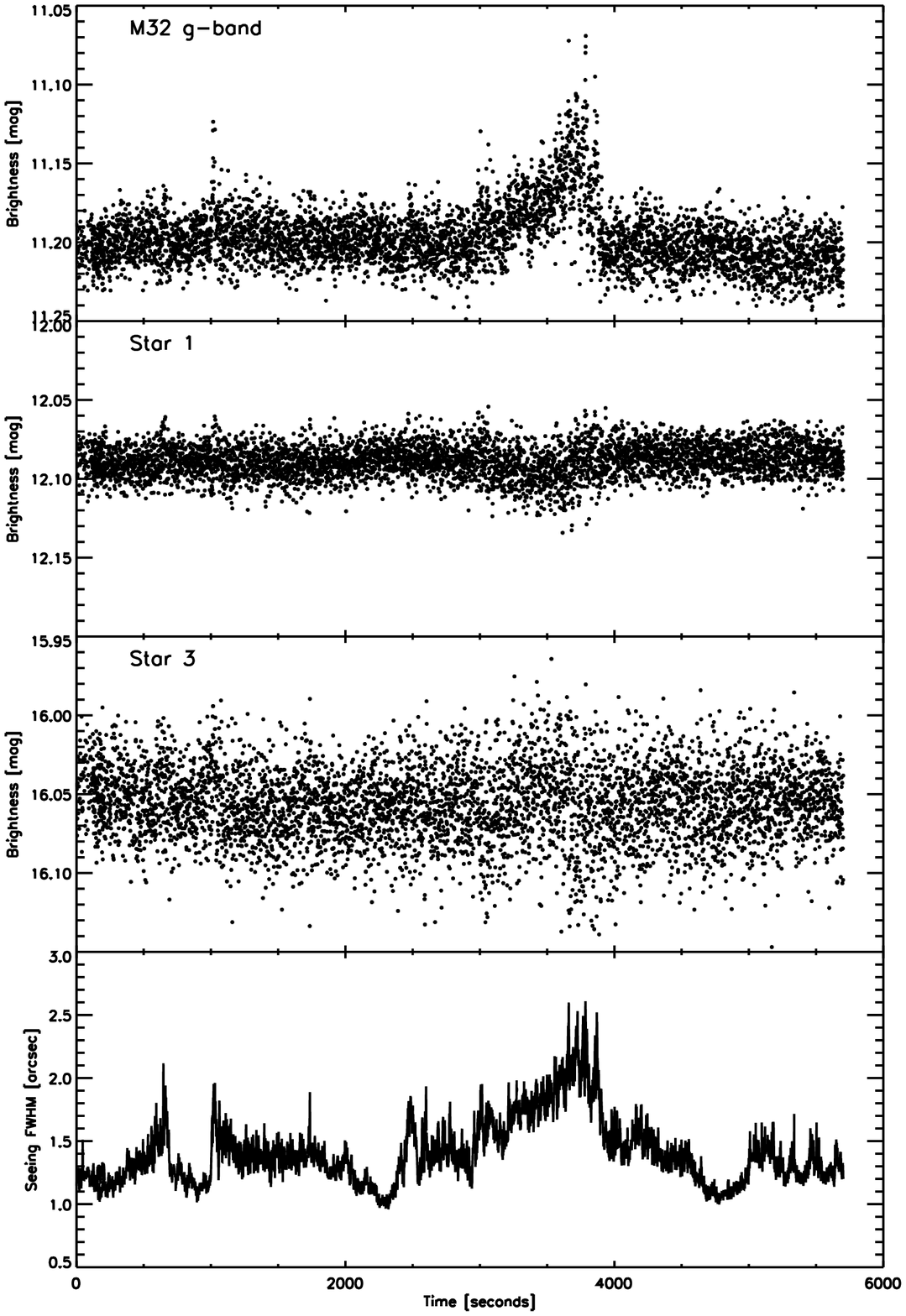}{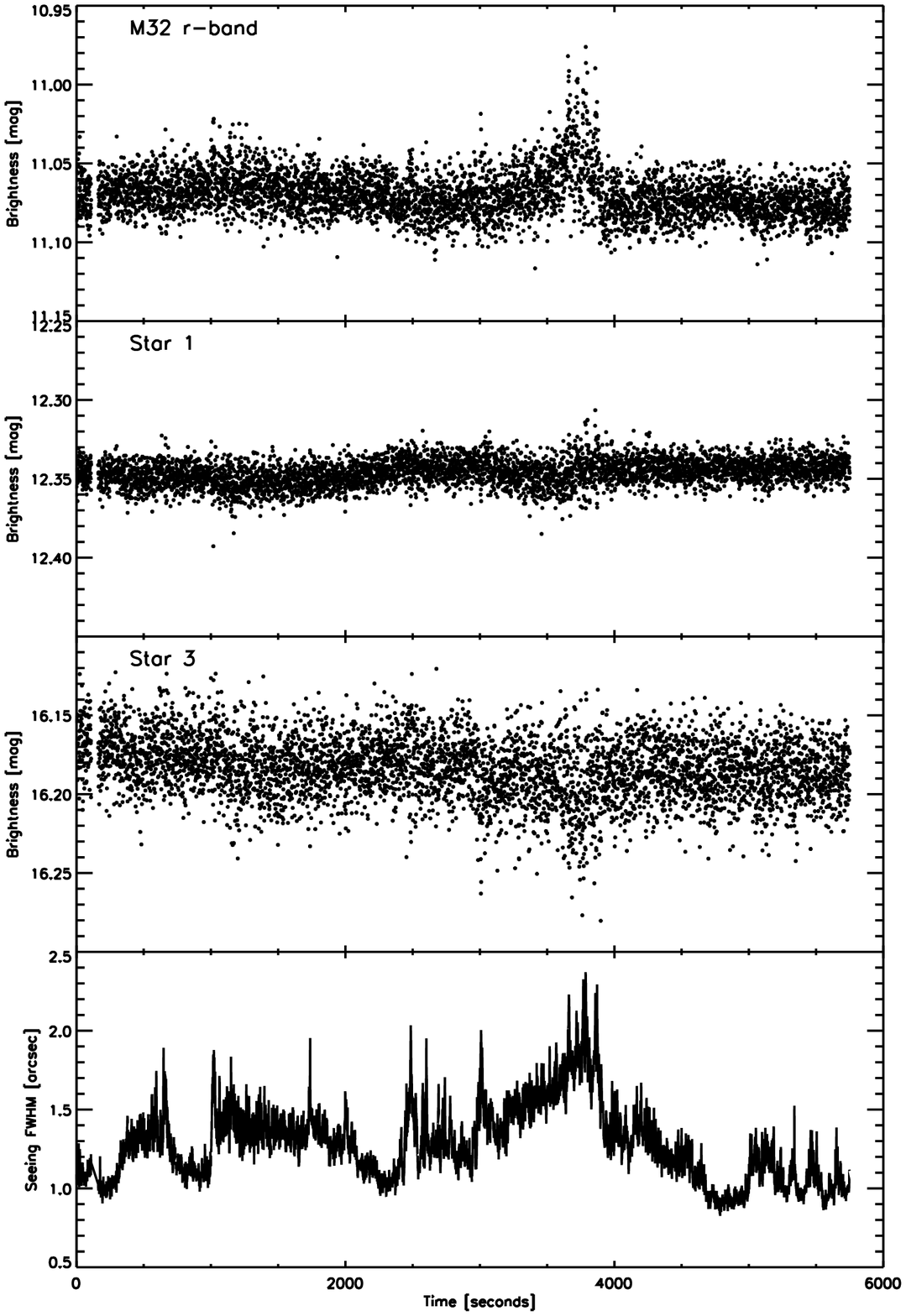}
\caption{Left: Calibrated $g'-$band light curve of the nucleus of the galaxy and two reference stars
after relative photometry with
respect to another reference star (Table 1). In the top
three panels, the abscissa shows time in seconds while the ordinate shows photometric brightness in mags, spanning a range of only 200 millimags. 
Clearly, photometric variations seen in figure \ref{fig1} have been dramatically reduced to the 1$\sigma\sim$10 millimag level through
relative photometry. The lower panel 
shows the variation of the seeing as determined from measurements of the full width at half maximum
(FWHM) of stars in the field. There clearly remains an additional term in the photometry
that is correlated with seeing; worse seeing causes a larger fraction of bulge light to enter the
the extraction aperture causing the M32 nuclear brightness to increase while the reference star brightness marginally decreases. Times of bad seeing are masked out in the
time series analysis as described in Section 2. Right: Corresponding calibrated $r'-$band light curve as on the left.}
\label{fig2}
\end{center}
\end{figure}

\begin{figure}[htbp]
\begin{center}
\plotone{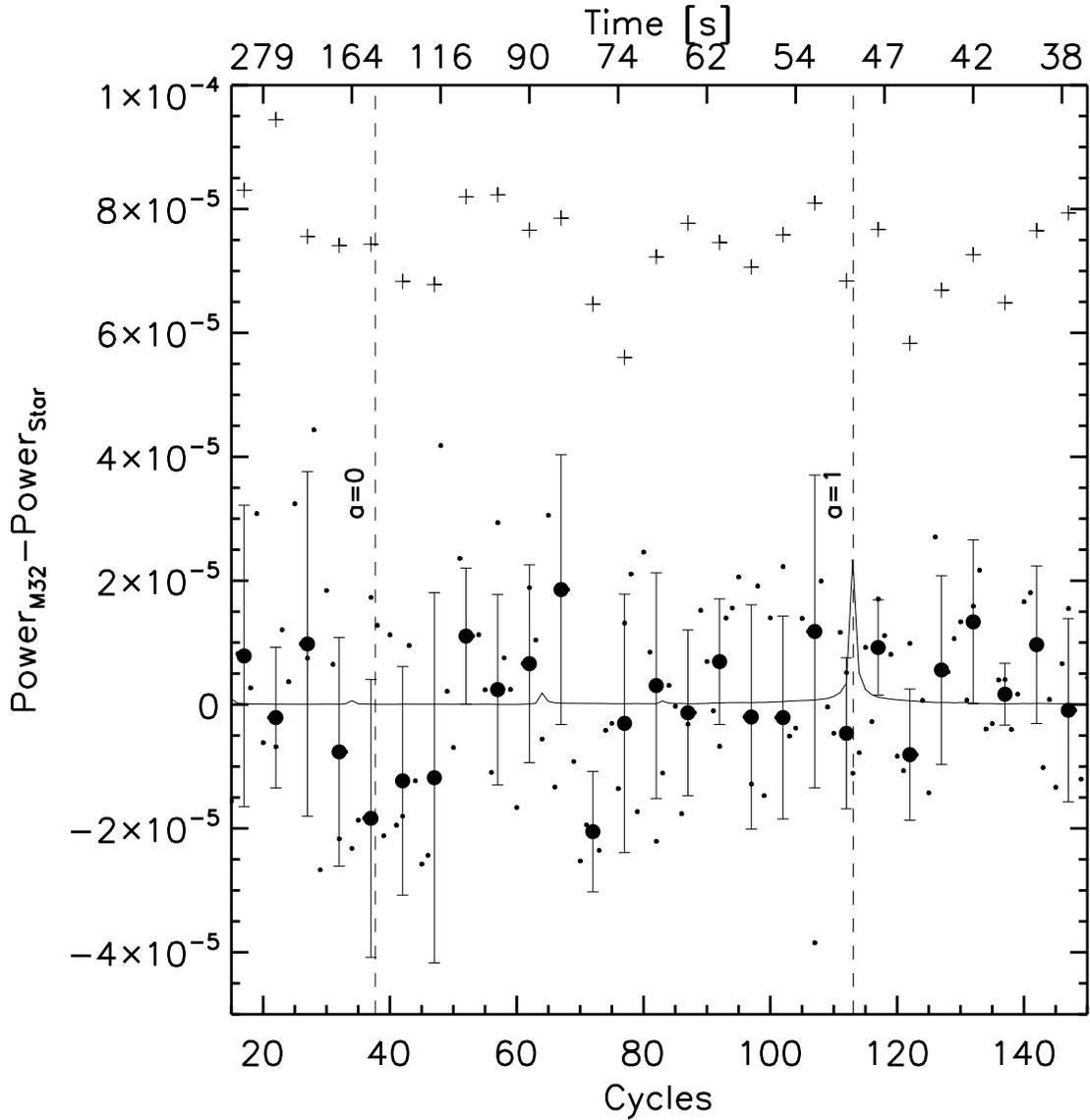}
\caption{
Binned excess in the temporal power spectrum in the $g'-$band for M32 (solid circles). The same difference for a reference star is shown
as plus symbols and has been offset by 7$\times10^{-5}$ for clarity.
The abscissa is the number of cycles over the duration of our observations which is the inverse of the time axis (top). The unbinned points for M32 are shown as the small dots.
There is no evidence for excess power in the power spectrum which might arise from optical flaring in the nucleus.
By taking the difference between the power spectrum from calibrated light curves, most observational systematics have been eliminated. 
Only times when seeing was better than 1.5$\arcsec$ was considered. 
The vertical dashed lines indicate the light travel time of the innermost circular stable orbit (ISCO) corresponding to a non-spinning ($a=0$) or a maximally spinning ($a=1$) black hole with mass of 2.5$\times10^{6}$\,M$_{\sun}$.
Also shown as a solid line is the power spectrum for a sinusoid
of period 49\,s, the light travel time corresponding to the ISCO for a maximally spinning black hole.
}
\label{fig3}
\end{center}
\end{figure}

\clearpage
\begin{deluxetable}{rccccl}
\tablecaption{Stars used for relative photometry}
\tablehead{
\colhead{$(\alpha,\delta)$ in J2000} &&&&& \colhead{{\it Gaia} mag}
}
\startdata
00:42:53.40, +40:52:15.91 & Star 1 &&&& 12.309\\
00:42:49.76, +40:52:11.30 & Star 2 &&&& 14.968 \\
00:42:34.07, +40:53:10.95 & Star 3 &&&& 16.133 \\
00:42:36.21, +40:49:31.75 & Star 4 &&&& 15.591 \\
\enddata
\label{tbl1}
\end{deluxetable}

\begin{deluxetable}{rccccccccc}
\tablecaption{Scatter in $g'-$band light curve of a star using different techniques}
\tablehead{
\colhead{Technique} &&&&&&&& \multicolumn{2}{c}{$\sigma_{\rm milli-mag}$}\\ 
\cline{9-10}
\colhead{} &&&&&&&& \colhead{12th mag} & \colhead{16th mag}
}
\startdata
Two dimensional Gaussian fits                       &&&&&&&& 61 & 60 \\
Fixed aperture photometry with local sky       &&&&&&&& 10 & 23\\
Fixed aperture photometry with common sky &&&&&&&& 11 & 30 \\
Variable aperture photometry with local sky   &&&&&&&& 13 & 26
\enddata
\label{tbl2}
\end{deluxetable}

\end{document}